\documentclass[12pt]{iopart}
\usepackage{amssymb}
\usepackage{graphicx}

\begin{document}

\title{Thermodynamic fluctuation relation for temperature and energy}
\author{L. Velazquez$^{1,2}$ and S. Curilef$^{2,3}$}

\begin{abstract}
The present work extends the well-known thermodynamic relation $C=\beta
^{2}\left\langle \delta {E^{2}}\right\rangle $ for the canonical ensemble.
We start from the general situation of the thermodynamic equilibrium between
a large but finite system of interest and a generalized thermostat, which we
define in the course of the paper. The resulting identity $\left\langle
\delta \beta \delta {E}\right\rangle =1+\left\langle \delta {E^{2}}%
\right\rangle \partial ^{2}S\left( E\right) /\partial {E^{2}}$ can account
for thermodynamic states with a negative heat capacity $C<0$; at the same
time, it represents a thermodynamic fluctuation relation that imposes some
restrictions on the determination of the microcanonical caloric curve $\beta
\left( E\right) =\partial S\left( E\right) /\partial E$. Finally, we comment
briefly on the implications of the present result for the development of new
Monte Carlo methods and an apparent analogy with quantum mechanics.\newline
\newline
PACS: 05.20.Gg; 05.40.-a; 75.40.-s\newline
\emph{Keywords}: Classical Ensemble Theory; Fluctuation Theory; Specific
Heats
\end{abstract}

\address{$^1$Departamento de F\'\i sica, Universidad de Pinar del R\'\i o,
Mart\'\i \,\,270, Esq. 27 de Noviembre, Pinar del R\'\i o, Cuba.
\\$^2$Departamento de F\'\i sica, Universidad Cat\'olica del Norte, Av. Angamos 0610, Antofagasta,
Chile.\\$^3$ Physics Department, University of Pretoria, Pretoria 0002, South Africa.}

\section{Introduction}

From the standard perspective of Statistical Mechanics, it is costumary to
start with the Gibbs canonical ensemble given by:
\begin{equation}
\hat{\omega}_{c}\left( \beta ,N\right) =\frac{1}{Z\left( \beta ,N\right) }%
\exp \left\{ -\beta \hat{H}_{N}\right\} ,  \label{gibbs}
\end{equation}%
which provides the macroscopic description of a Hamiltonian system $\hat{H}%
_{N}$ in thermodynamic equilibrium with a heat bath (a very large heat
reservoir) at constant temperature $T$, where $\beta =1/k_{B}T$. Hereafter,
the Boltzmann constant $k_{B}$ is set to 1. In this context, other
thermodynamic parameters of the system, like the volume $V$ or an external
magnetic field $H$, are also admissible, but\ we assume along this paper
that every parameter remains constant. A straightforward consequence of
using this kind of statistical ensemble is the relation between the heat
capacity $C=\partial E/\partial T$ and the canonical average of the energy
fluctuation $\left\langle \delta E^{2}\right\rangle $:
\begin{equation}
C=\beta ^{2}\left\langle \delta E^{2}\right\rangle ,  \label{fluct1}
\end{equation}%
which leads to the \textit{non-negative} character of the heat capacity
within the canonical description, e.g. $C_{V}>0$ for a fluid with volume $V$%
, or $C_{H}>0$ for a magnetic system in external magnetic field $H$. This
same result can be also derived from the stability analysis in the framework
of the standard Thermodynamics, therefore, it is usually claimed in several
classical textbooks that those macrostates, in which this condition is not
satisfied, are thermodynamically unstable and \textit{cannot exist in Nature}
(see, for instance, in the section \S 21 of the Landau \& Lifshitz book \cite%
{Landau}).

Surprisingly, macrostates with negative heat capacities have been \textit{%
actually observed} in several systems belonging to different physical
scenarios. For example, in the mesoscopic short-range interacting systems
like the nuclear, atomic and molecular clusters \cite%
{moretto,Ison,Dagostino,gro na}, as well as the long-range interacting
systems like astrophysical ones \cite{Lynden,LyndenW,einarsson}, all of them
often referred as \textit{nonextensive systems}\footnote{%
Roughly speaking, a system is nonextensive when it cannot be trivially
divided in independent subsystems, which is explained by the existence of
underlying interactions or correlation effects whose characteristic length
is comparable or larger than the system linear size. Thus, the total energy
in such systems is nonadditive, and frequently, they are spatially
nonhomogeneous.}. This observation illustrates the limited validity of
certain standard results of classical Thermodynamics and Statistical
Mechanics in such contexts \cite{Dauxois}. We are going to provide two
illuminating examples below, but first, let us explain how negative heat
capacities arise in the thermodynamic description.

The usual way to access macrostates with negative heat capacities is by
means of the microcanonical description. The fundamental key is to rephrase
the heat capacity $C=\partial E/\partial T$ in terms of the Boltzmann
entropy $S=\ln {W}$. Starting from the definition of the microcanonical
inverse temperature of the system:%
\begin{equation}
\frac{\partial S}{\partial E}=\frac{1}{T},
\end{equation}%
we obtain then the second derivative of the entropy:
\begin{equation}
\frac{\partial ^{2}S}{\partial E^{2}}=-\frac{1}{T^{2}}\frac{\partial T}{%
\partial E}\Rightarrow C=-\left( \frac{\partial S}{\partial E}\right)
^{2}\left( \frac{\partial ^{2}S}{\partial E^{2}}\right) ^{-1}.
\label{micro.cap}
\end{equation}%
Since the Boltzmann entropy is a geometrical measure of the microcanonical
ensemble, it demands neither the extensive and concave properties usually
attributed to its probabilistic interpretation:
\begin{equation}
S_{e}\left[ p\right] =-\sum_{k}p_{k}\ln p_{k},
\end{equation}%
nor the consideration of the thermodynamic limit. Thus, the expression (\ref%
{micro.cap}) states that negative heat capacities are directly related to
the presence of macrostates with a \textit{convex} microcanonical entropy, $%
\partial ^{2}S/\partial E^{2}>0$.

\begin{figure}[tbp]
\begin{center}
\includegraphics[width=4.0413in]{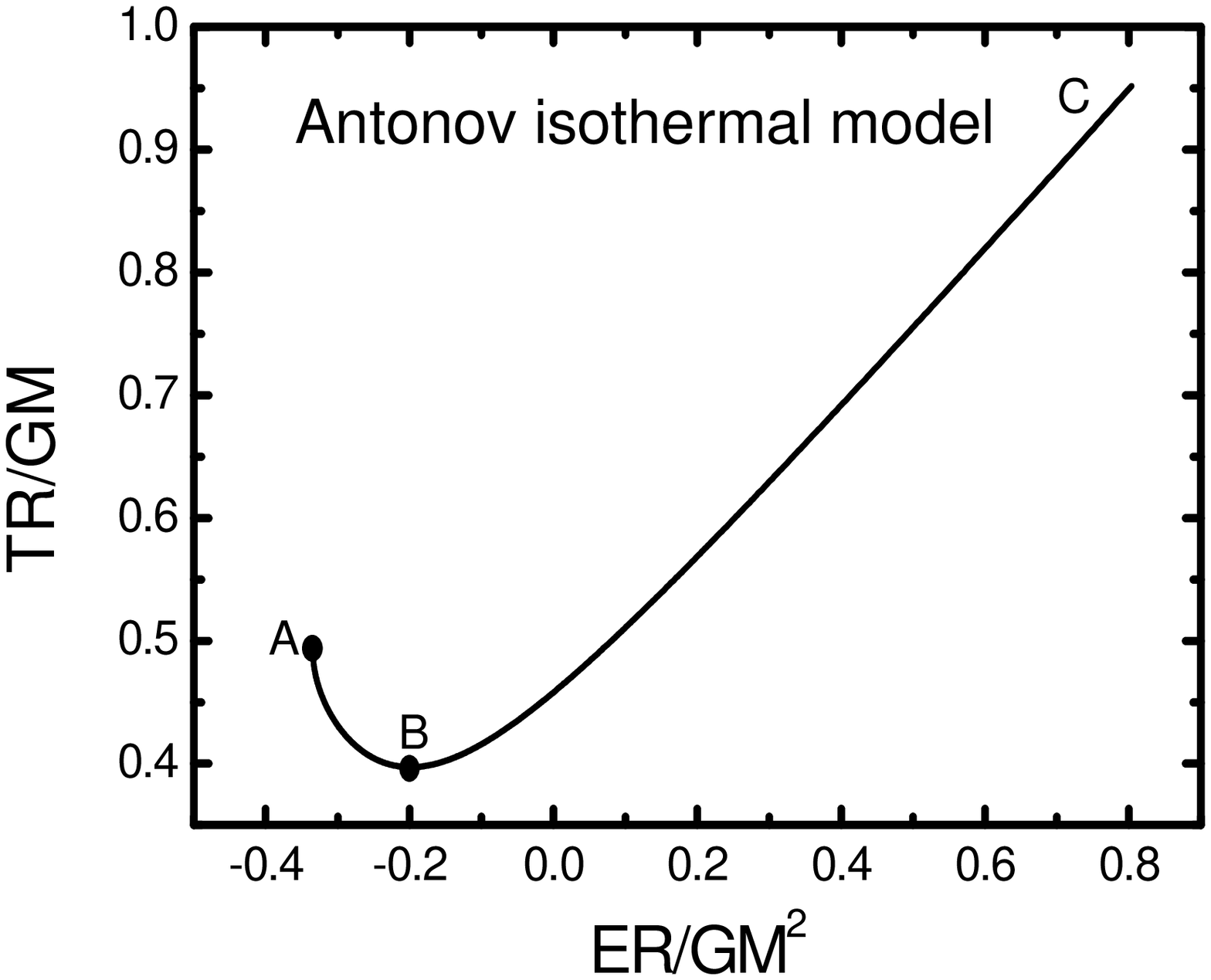}
\end{center}
\caption{Microcanonical caloric curve of the Antonov isothermal model (after
\protect\cite{antonov}). The branch \textbf{AB} represents macrostates where
the heat capacity $C_{V}<0$. Notice also the nonexistence of equilibrium
macrostates when $T<T_{B}\simeq 0.4GM/R$ or $E<E_{A}\simeq -0.335GM^{2}/R$,
which is related to the occurrence of a gravitational collapse at low
temperatures (canonical ensemble) or at low energies (microcanonical
ensemble).}
\label{antonov.eps}
\end{figure}

To our knowledge, the existence of negative heat capacities was firstly
pointed out by Lynden-Bell in the astrophysical context in seminal papers
\cite{Lynden,LyndenW}. Interestingly, the presence of this anomaly plays a
fundamental role in understanding the evolution of these remarkable physical
systems \cite{thyr}. A simple astrophysical model that shows an energetic
region with $C_{V}<0$ is the Antonov isothermal model \cite{antonov}: a
system of self-gravitating identical point particles with a total mass $M$
enclosed in a rigid spherical container of radius $R$, whose microcanonical
caloric curve is depicted in FIG.\ref{antonov.eps}.

\begin{figure}[tbp]
\begin{center}
\includegraphics[width=4.0413in]{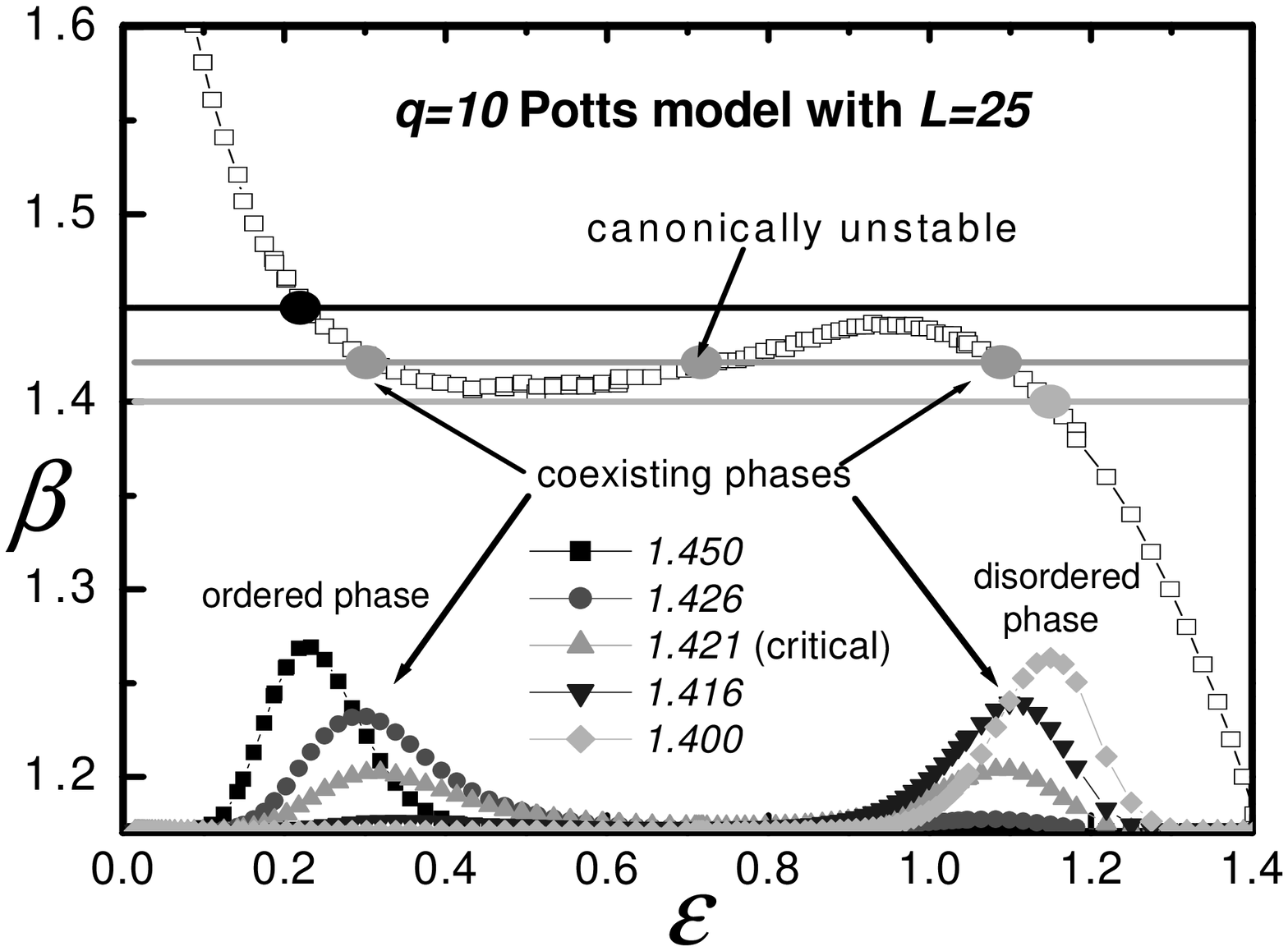}
\end{center}
\caption{The energy distribution functions within the canonical ensemble,
obtained from Eq.(\protect\ref{cedf}) by using $\Omega \left( E\right) $
estimated from Monte Carlo calculations (see sec. \protect\ref{control.m}
below), indicate the occurrence of phase coexistence phenomenon in the $q=10$
states Potts model on the square lattice $N=L\times L$ with periodic
boundary conditions ($\protect\varepsilon =E/N$). Notice that the
microcanonical states with negative heat capacities are practically
inaccessible by using a thermostat with constant temperature, since these
macrostates are canonically unstable.}
\label{potts.eps}
\end{figure}

The existence of macrostates with negative heat capacities can be also
related to the occurrence of phase coexistence phenomenon or first-order
transitions in finite short-range interacting systems prior to any
thermodynamic limit \cite{Lyn2,pad,Lyn3,gro1}. Such a relationship is
clearly illustrated in FIG.\ref{potts.eps} for the $q=10$ states Potts model
on a square lattice $N=L\times L$:%
\begin{equation}
H_{N}=\sum_{\left\langle i,j\right\rangle }\left( 1-\delta _{\sigma
_{i},\sigma _{j}}\right) ,  \label{potts}
\end{equation}%
with periodic boundary conditions, where the sum considers only the nearest
neighbor interactions \cite{wolf}. Here, the microcanonical caloric curve $%
\partial S/\partial E$ in terms of the energy per particle $\varepsilon =E/N$
shows a backbending indicating the presence of an anomalous region where $%
C_{H}<0$ at $H=\left\vert \mathbf{H}\right\vert =0$\footnote{%
The case with $\left\vert \mathbf{H}\right\vert \not=0$ is described by the
Hamiltonian $\mathcal{H}_{N}=\hat{H}_{N}-\mathbf{H\cdot M}$, where the total
magnetization $\mathbf{M}=\sum_{i}\mathbf{s}_{i}$ and $\mathbf{s}_{i}=\left[
\cos \left( \kappa \sigma _{i}\right) ,\sin \left( \kappa \sigma _{i}\right) %
\right] $ with $\kappa =2\pi /q$.}. In the neighborhood of the critical
point $\beta _{c}\simeq 1.42$, the bimodal character of the energy
distribution function within the canonical ensemble:%
\begin{equation}
\rho _{c}\left( E\right) dE=\frac{1}{Z\left( \beta \right) }\exp \left(
-\beta E\right) \Omega \left( E\right) dE,  \label{cedf}
\end{equation}%
($\Omega \left( E\right) $ is the state density of the system) reveals
\textit{two coexisting phases} with different energies at the same
temperature (ferromagnetic and paramagnetic phases). The localization of the
peaks is determined from those intersection points of the microcanonical
caloric curve $\partial S/\partial E$ where $\partial ^{2}S/\partial E^{2}<0$
with the horizontal lines representing here the inverse temperature $\beta $
of the thermostat (the ordinary equilibrium condition between the thermostat
and the system temperature). Since no one peak accesses the region where $%
\partial ^{2}S/\partial E^{2}>0$ at any thermostat temperature, macrostates
with $C_{H}<0$ are \textit{practically inaccessible} (unstable)\ within the
canonical ensemble when the system of interest is large enough. Such a
"forbidden" region is the origin of the sudden jump of the canonical
average\ energy $\left\langle E\right\rangle $ at the neighborhood of the
critical point $\beta _{c}$, which tell us about the existence of a latent
heat $q_{L}$ for the conversion of one phase into another.

The close relation between macrostates with negative heat capacities and
first-order phase transitions in finite short-range interacting systems
clarifies that such an "anomalous" behavior is far from an unusual feature
within the thermostatistical description, including also all those systems
which have been traditionally considered within the standard Thermodynamics%
\footnote{%
The consideration of the thermodynamic limit $N\rightarrow \infty $ with $%
N/V $ fixed in the thermo-statistical description of short-range interacting
systems is a useful and convenient \ idealization which dismisses the
occurrence of boundary effects. In practice, any physical system is conforms
by a very large but finite number of constituents. In fact, the existence of
backbending of the microcanonical caloric curve as in FIG.\ref{potts.eps} is
just a finite \textit{size effect} associated to the presence of interphases
during the occurrence of the first-order phase transitions (see in \cite%
{gro1}).}. The anomalous character of such macrostates reflects that they
are physically admissible within the microcanonical description, while they
are thermodynamically unstable within the canonical description; indicating
thus the\textit{\ inequivalence} of these statistical ensembles for finite
systems.

As already evidenced, macrostates with $C<0$ do not receive a correct
treatment by the fluctuation relations derived from the usual equilibrium
situations of the standard Statistical Mechanics and Thermodynamics: in
fact, the condition $C<0$ in Eq.(\ref{fluct1}) cannot be realized in a
canonical description, where $\beta =$ constant, or within a microcanonical
framework, where $E=$ constant so $\delta E$ vanishes. Precisely, the
fundamental aim of this work is to obtain a suitable generalization of
thermodynamic relation (\ref{fluct1}) in which accounts for appropriately
the existence of macrostates with $C<0$.

The generalized fluctuation relation resulting from the present analysis
points out the role of macrostates with $C<0$ in the experimental
determination of the microcanonical caloric curve (which implies a
simultaneous measurement of the temperature and energy of a given system),
contributing in this way to a re-examination of an old question of the
Thermo-Statistics Theory: \textit{Could it be possible the existence of
certain kind of complementarity between the energy and the temperature}?
Such an idea was suggested by Bohr and Heisenberg in the early days of the
Quantum Mechanics \cite{bohr}, and so far, it have not received a general
consensus in the scientific literature \cite{Uffink}.

This paper is organized into sections, as follows: First, in section 2, we
introduce concepts like the generalized thermostat, effective inverse
temperature and fluctuation in the Gaussian approximation to derive our
fundamental result, Eq.(\ref{tur}). Second, in section 3, we discuss some
implications on themodynamic control and measurements. Furthermore, \ we
perform a generalization of this result, Eq.(\ref{def.tur}), which
constitutes an "uncertainty relation" in Thermodynamics. Finally, in section
4, we make some concluding remarks about several theoretical and practical
implications of the present formalism.

\section{Extending fluctuation relation}

\subsection{Generalized thermostat}

Derivation of a generalized fluctuation relation (\ref{fluct1}) to deal with
macrostates with $C<0$ demands to face the problem of the ensemble
inequivalence, to do this, we need to start from a more general equilibrium
situation of the kind "system-surrounding", where macrostates with negative
heat capacities can be arisen as thermodynamically stable under the external
influence imposed by the system surrounding.

In thermodynamic equilibrium, the underlying physical conditions of the
natural environment lead to a situation that a given short-range interacting
system obeys the Gibbs canonical ensemble (\ref{gibbs}). The natural
environment is a good example of a Gibbs thermostat, since its heat capacity
is so large that it can be considered to be infinite in every practical
situation. According to Lynden-Bell in ref.\cite{Lyn3}, a system with a
negative heat capacity $C_{1}<0$ can reach thermodynamic equilibrium with a
second system with $C_{2}>0$ when $0<C_{2}<\left\vert C_{1}\right\vert $.
Obviously, such a condition does not hold when the first system is under the
influence of the natural environment, since $C_{2}\rightarrow \infty $, and
therefore, it is not possible to access the macrostates with negative heat
capacities. As a corollary of this reasoning, the direct observation of the
macrostates with negative heat capacities of a given system demands that the
external influence of the natural environment be suppressed. Thus, the
system of interest can be isolated (microcanonical ensemble) or it can be
under the influence of a second system that remains of finite size. The
latter possibility allows to express the energy distribution function of the
first subsystem as follows:
\begin{equation}
\rho \left( E_{1}\right) dE_{1}=\frac{1}{\Omega _{c}\left( E_{T}\right) }%
\Omega _{2}\left( E_{T}-E_{1}\right) \Omega _{1}\left( E_{1}\right) dE_{1},
\label{finite}
\end{equation}%
where $\Omega _{c}\left( E_{T}\right) =\int_{0}^{E_{T}}\Omega _{2}\left(
E_{T}-E_{1}\right) \Omega _{1}\left( E_{1}\right) dE_{1}$ and as before, $%
\Omega _{2}\left( E_{2}\right) $ is the state density of the second system
acting as a "finite thermostat", and $E_{T}$ is the total energy of the
closed system $E_{T}=E_{1}+E_{2}$.

After reading the analysis presented in the subsequent subsections, it can
be realized that the consideration of the ansatz (\ref{finite}) is
sufficient to arrive at a generalized expression of the fluctuation relation
(\ref{fluct1}): a closed system composed of two independent \textit{finite}
subsystems with an \textit{additive} total energy. However, such a physical
picture is only admissible when these subsystems are coupled by the
incidence of short-range interacting forces or when long-range forces are
confined to each subsystem which, however, is in general nonphysical. Even
in this case this assumption presupposes to dismiss the energy contribution
involved in their mutual interactions $V_{int}$. Although this is a licit
and useful approximation in standard applications of Statistical Mechanics
and Thermodynamics, it may be unrealistic in the case of mesoscopic systems
with short-range interactions, and worse, this approach is not applicable in
the case of long-range interacting systems. The latter case constitutes a
typical scenario where macrostates with negative heat capacities naturally
appear. Obviously, the additivity of the total energy is no longer
applicable since the interaction energy $V_{int}$ cannot be considered as a
"boundary effect" as in the case of large systems with short-range
interactions, and often, separability of a closed system into several parts
is a hypothesis that should be carefully applied \cite{land}.

Nevertheless, we can find some systems in Nature where it is still possible
to assume certain separability of a closed system into subsystems despite
the presence of long-range interactions. Good examples are galaxies and
their clusters. Of course, we are unable to dismiss the interactional energy
$V_{int}$ in this scenario, but it is reasonable to assume as a consequence
of the separability that this energy contribution could be approximated by
certain functional dependence of the internal energies of each subsystems, $%
V_{int}\simeq V_{int}\left( E_{1},E_{2}\right) $. Thus, the usual additivity
of the total energy $E_{T}$\ could be substituted by the following ansatz:
\begin{equation}
E_{T}=\Phi \left( E_{1},E_{2}\right) \simeq E_{1}+E_{2}+V_{int}\left(
E_{1},E_{2}\right) .  \label{non.additive}
\end{equation}%
The number of macrostates of the whole system $\Omega _{c}\left(
E_{T}\right) $ is given by:
\begin{equation}
\Omega _{c}\left( E_{T}\right) =\int \delta \left[ E_{T}-\Phi \left(
E_{1},E_{2}\right) \right] \Omega _{1}\left( E_{1}\right) \Omega _{2}\left(
E_{2}\right) dE_{1}dE_{2},
\end{equation}%
which allows to express the energy distribution of the first subsystem as
follows:%
\begin{equation}
\rho \left( E_{1};E_{T}\right) dE_{1}=\omega \left( E_{1};E_{T}\right)
\Omega _{1}\left( E_{1}\right) dE_{1},  \label{rel2}
\end{equation}%
and the probabilistic weight $\omega \left( E_{1};E_{T}\right) $ by:%
\begin{equation}
\omega \left( E_{1};E_{T}\right) =\frac{1}{\Omega _{c}\left( E_{T}\right) }%
\int \delta \left[ E_{T}-\Phi \left( E_{1},E_{2}\right) \right] \Omega
_{2}\left( E_{2}\right) dE_{2}.  \label{prob.w.nl}
\end{equation}%
Eq.(\ref{rel2}) constitutes a more general expression than Eq.(\ref{finite}%
), providing a better treatment for a \textit{nonlinear energy interchange}
between the subsystems as a consequence of the nonadditivity of the total
energy ($\Delta E_{1}\not=-\Delta E_{2}$). This kind of consideration could
be applicable to both: the case of large systems with long-range
interactions and mesoscopic systems with short-range interactions.

Previous discussions have suggested that a general way to account for the
energy distribution function $\rho \left( E\right) dE$, which is associated
to a general "system-surrounding" equilibrium situation, is provided by the
ansatz:%
\begin{equation}
\rho \left( E\right) dE=\omega \left( E\right) \Omega \left( E\right) dE,
\label{new}
\end{equation}%
where $\Omega \left( E\right) $ represents the state density of the system
with internal energy $E$, and $\omega \left( E\right) $, a generic
probabilistic weight characterizing the energetic interchange of this system
with its surrounding. The above hypothesis is very economical since it
demands merely: (1) the existence of some kind of separability between the
system and its surrounding, (2) and those all external influences on the
system are \textit{fully described} by the probabilistic weight $\omega
\left( E\right) $. In this work, we are admitting the validity of Eq.(\ref%
{new}) without mattering the internal structure of the surrounding, and
even, the \textit{features of its internal equilibrium conditions}. This
last idea is very important and deserves to be clarified.

It is almost a rule that a large system with long-range interactions,
initially far from the equilibrium, reaches rapidly a \textit{metastable
equilibrium} spending a long time in it \cite{Dauxois}. If this is the case,
the energy interchange of this large system acting as "surrounding" of
another system cannot be dealt by the expressions (\ref{finite}) or (\ref%
{rel2}). However, this metastability \textit{does not forbid} the
applicability of the ansatz (\ref{new}) in many physical situations. For
example, it is well known that the collisionless dynamics of astrophysical
systems leads to a metastable state where the one-particle distribution
function $f\left( q,p\right) $ depends only on the particle energy $%
\varepsilon \left( q,p\right) =\frac{1}{2m}p^{2}+m\varphi \left( q\right) $%
,\ where $f\left( q,p\right) =f\left[ \varepsilon \left( q,p\right) \right] $%
, whose mathematical form is not Boltzmannian and it is determined from the
initial conditions of dynamics \cite{inChava}. Since the macroscopic
behavior of the system in this metastable state is also ruled by its energy,
we can expect that this physical quantity also rules its interaction with
other systems. The admittance of the validity of the ansatz (\ref{new}) for
systems in metastable conditions accounts for the claims of some recent
authors about the existence of non-Boltzmannian energy distribution
functions outside the equilibrium, overall, in systems with a complex
microscopic dynamics, e.g.: astrophysical systems \cite{inChava} and
turbulent fluids \cite{beck1}. A unifying framework for many of \ these
distribution functions is provided by the so-called "Superstatistics", a
theory recently proposed by C. Beck and E.G.D Cohen \cite{cohen}, where the
presence of a non-Boltzmannian weight $\omega \left( E\right) $ seems to be
originated from an effective incidence of a fluctuating inverse temperature $%
\beta $ at the microscopic level (e.g. on a Brownian particle) obeying the
distribution function $f\left( \beta \right) $ as follows:%
\begin{equation}
\omega \left( E\right) =\int \exp \left( -\beta E\right) f\left( \beta
\right) d\beta .  \label{b.super}
\end{equation}

Since the microscopic origin and the specific mathematical form of the
probabilistic weight $\omega \left( E\right) $ are arbitrary, it is expected
that \textit{the conclusions, which are derived from the ansatz} (\ref{new}%
), \textit{may be applicable to a wide range of equilibrium or
meta-equilibrium situations}. In this formula, $\omega \left( E\right) $
constitutes a generic extension of the usual canonical weight:%
\begin{equation}
\omega _{c}\left( E;\beta \right) =C\exp \left( -\beta E\right) ,
\label{w.can}
\end{equation}%
which rules the energy interchange of a Gibbs thermostat (a very large
short-range interacting system). In an analogous way, we shall hereafter
refer to the system surrounding associated with the weight $\omega \left(
E\right) $ as a \textit{generalized thermostat}. Let us now show that this
generalized thermostat can be also characterized by certain \textit{%
effective inverse temperature} $\beta _{\omega }$, which controls, as usual,
the energetic interchange of this thermostat with the system under study.

\subsection{Effective inverse temperature $\protect\beta _{\protect\omega }$}

A straightforward way to arrive at this important thermodynamic quantity is
by using the probabilistic weight $\omega \left( E\right) $ in the
Metropolis Monte Carlo simulation of the system dynamics in this equilibrium
condition. The acceptance probability for a Metropolis move $p\left( \left.
E\right\vert E+\Delta E\right) $ is given by:%
\begin{equation}
p\left( \left. E\right\vert E+\Delta E\right) =\min \left\{ \frac{\omega
\left( E+\Delta E\right) }{\omega \left( E\right) },1\right\} .  \label{p1}
\end{equation}%
By assuming that the system size is large enough, consequently, the amount
of energy $\left\vert \Delta E\right\vert \ll \left\vert E\right\vert $, we
are able to introduce the approximation:
\begin{equation}
\ln \frac{\omega \left( E+\Delta E\right) }{\omega \left( E\right) }\simeq
\frac{\partial \ln \omega \left( E\right) }{\partial E}\Delta E,
\end{equation}%
and rephrase the acceptance probability (\ref{p1}) as follows:%
\begin{equation}
p\left( \left. E\right\vert E+\Delta E\right) \simeq \min \left\{ \exp \left[
-\beta _{\omega }\left( E\right) \Delta E\right] ,1\right\} .
\label{interchange}
\end{equation}%
The evident analogy of this last results with the canonical case allows to
define the quantity $\beta _{\omega }\left( E\right) $
\begin{equation}
\beta _{\omega }\left( E\right) =-\frac{\partial \ln \omega \left( E\right)
}{\partial E}  \label{effective.b}
\end{equation}%
as the effective inverse temperature of the generalized thermostat.

It is possible to verify that this definition is not arbitrary. Besides the
obvious case of the canonical weight (\ref{w.can}) where $\beta _{\omega
}\left( E\right) \ $is the inverse temperature of the Gibbs thermostat, $%
\beta _{\omega }\left( E\right) \equiv \beta $, the present definition drops
to the usual microcanonical inverse temperature of the system acting as a
"finite thermostat" in Eq.(\ref{finite}), where $\omega \left( E\right)
=\Omega _{2}\left( E_{2}\right) /\Omega \left( E_{T}\right) $ and $\beta
_{\omega }\left( E\right) \equiv \beta _{2}\left( E_{2}\right) =\partial
S_{2}\left( E_{2}\right) /\partial E_{2}$, being $S_{2}\left( E_{2}\right)
=\ln \left[ \Omega _{2}\left( E_{2}\right) \delta E_{0}\right] $ and $%
E_{2}=E_{T}-E$.

This inverse temperature is "effective" because it is not always equivalent
to the ordinary interpretation of this concept in the standard Statistical
Mechanics. In order to verify this fact, let us consider the equilibrium
situation accounted for by Eq.(\ref{rel2}). The internal energy of the
second subsystem $E_{2}$ can be expressed as follows:
\begin{equation}
E_{2}=E_{T}-E_{1}-V_{int}\left( E_{1},E_{2}\right) ,
\end{equation}%
where the recursive substitution of this last equation in $V_{int}\left(
E_{1},E_{2}\right) $ leads to the certain nonlinear dependence of $E_{2}$ on
the internal energy of the first subsystem $E_{1}$:
\begin{equation}
E_{2}=E_{2}\left( E_{1}\right) \equiv E_{T}-\Theta \left( E_{1};E_{T}\right)
.
\end{equation}%
Thus, the probabilistic weight of Eq.(\ref{prob.w.nl}) is similar to the
case of the "finite thermostat":%
\begin{equation}
\omega \left( E_{1};E_{T}\right) =\frac{\Omega _{2}\left[ E_{2}\left(
E_{1}\right) \right] }{\Omega \left( E\right) },
\end{equation}%
but the effective inverse temperature $\beta _{\omega }$ is given by:%
\begin{equation}
\beta _{\omega }\left( E_{1}\right) =\frac{\partial S_{2}\left( E_{2}\right)
}{\partial E_{2}}\nu \left( E_{1}\right) ,  \label{beta.non.additive}
\end{equation}%
where the factor $\nu \left( E_{1}\right) =-\partial E_{2}/\partial
E_{1}=\partial \Theta \left( E_{1};E_{T}\right) /\partial E_{1}\not=1$
accounts for the existence of a nonlinear energy interchange as a
consequence of the nonadditivity of the total energy. It is remarkable that
the "effective inverse temperature" of a very large system surrounding
depends on the internal energy $E_{1}$ despite its "microcanonical inverse
temperature" $\beta _{2}=\partial S_{2}/\partial E_{2}$ remains practically
unaltered by the underlying energy interchange, as in the case of the Gibbs
thermostat.

The physical meaning of the effective inverse temperature $\beta _{\omega }$
is even more unclear in the case where the probabilistic weight $\omega
\left( E\right) $ is associated with a system surrounding in a metastable
equilibrium. Obviously, this kind of inverse temperature has nothing to do
with the inverse temperature $\beta $ of the integral representation (\ref%
{b.super}) of the Superstatistics formalism: while a whole set of values of $%
\beta $ for each energy $E$ exists, there is only one value of $\beta
_{\omega }$ for each value of $E$. The importance of this new concept relies
on the possibility to consider a wide class of equilibrium or
meta-equilibrium situations in a unifying framework where it could be
possible to extend the validity of some known thermostatistical results.

A fundamental identity is the condition of \textit{thermal equilibrium}.
This condition commonly follows from the analysis of \ the most probable
macrostate $\bar{E}$:%
\begin{equation}
\max_{\bar{E}}\left\{ \omega \left( E\right) \Omega \left( E\right) \right\}
\end{equation}%
which leads directly to the stationary equation:
\begin{equation}
\beta _{\omega }\left( \bar{E}\right) =\beta _{s}\left( \bar{E}\right) ,
\label{thermal.eq}
\end{equation}%
where $\beta _{s}\left( E\right) $ is just the microcanonical inverse
temperature of the system:
\begin{equation}
\beta _{s}\left( E\right) =\frac{\partial S\left( E\right) }{\partial E},
\end{equation}%
where, as in the introductory section, we take $S\left( E\right) =\ln W(E)$.
We have demonstrated that the quantity $\beta _{\omega }$ also obeys the
ordinary form of the \textit{Zeroth Principle of Thermodynamics }. By
analyzing the expression of $\beta _{\omega }$ shown in Eq.(\ref%
{beta.non.additive}) associated with the equilibrium between two separable
subsystems with a nonadditive total energy, Eq.(\ref{non.additive}), we
notice that it is precisely $\beta _{\omega }$, and not $\beta _{2}$, which
becomes equal to microcanonical inverse temperature of the first subsystem $%
\beta _{1}$ during the thermal equilibrium, $\beta _{\omega }=\beta
_{1}\Rightarrow \beta _{2}\not=\beta _{1}$. This result is essentially the
same as the one presented by Johal in the ref.\cite{Johal}.

Let us now summarize the fundamental properties of the effective inverse
temperature $\beta _{\omega }$:

\begin{description}
\item[A. ] It characterizes the energy interchange "system-surrounding"
during the equilibrium, in accordance with Eq.(\ref{interchange}).

\item[B. ] This concept permits to extend the thermal equilibrium condition,
Eq.(\ref{thermal.eq}), to a wide class of physical situations.

\item[C. ] In general, this thermodynamical quantity depends on the internal
energy of the system $E$, $\beta _{\omega }\left( E\right) $.
\end{description}

Properties A and B correspond to the ordinary physical notion of
temperature. A new remarkable property is the general dependence of $\beta
_{\omega }$ on the internal energy $E$ \ of the system (the property C).
This fact indicates clearly that the underlying energy interchange not only
imposes fluctuations on the internal energy $E$, but also provokes the
existence of \textit{correlated fluctuations} between the internal energy $E$
and the effective inverse temperature of the thermostat $\beta _{\omega }$, $%
\left\langle \delta \beta _{\omega }\delta E\right\rangle \not=0$. \ This
simple property has been systematically disregarded by the use of the Gibbs
canonical ensemble in standard Statistical Mechanics. We shall show in the
next subsection that the property C is precisely the fundamental key for
arriving at a suitable generalization of the fluctuation relation (\ref%
{fluct1}).

\begin{figure}[tbp]
\begin{center}
\includegraphics[width=4.0075in]{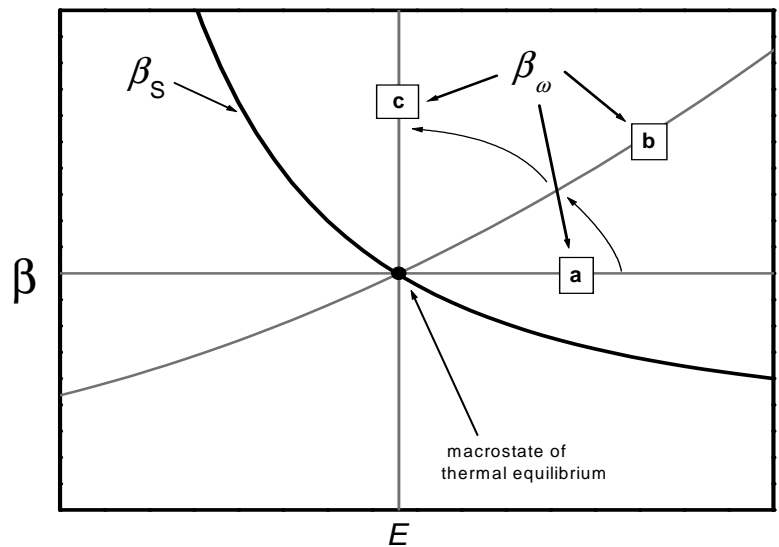}
\end{center}
\caption{Typical behavior of the effective inverse temperature $\protect%
\beta _{\protect\omega }$ is depicted for: a) $\protect\beta _{\protect%
\omega }$ \textit{fixed} for all values of $E$ (canonical ensemble), b) an
arbitrary dependence $\protect\beta _{\protect\omega }\left( E\right) $
(generalized thermostat), and c) $E$ \textit{fixed} for all values of $%
\protect\beta _{\protect\omega }$ (microcanonical ensemble). The black thick
point represents the macrostate of thermal equilibrium where $\protect\beta %
_{\protect\omega }=\protect\beta _{s}$, and $\protect\beta _{s}\left(
E\right) =\partial S\left( E\right) /\partial E$, the microcanonical inverse
temperature of the system.}
\label{gibbs.thermo.eps}
\end{figure}

Before the end of this subsection, it is important to remark that the
probabilistic weight:%
\begin{equation}
\omega _{m}\left( E;E_{0}\right) =C\delta \left( E-E_{0}\right) ,
\label{micro}
\end{equation}%
which is associated with the energetic isolation or the microcanonical
ensemble, can be also considered as a limiting case of a generalized
thermostat. The application of the definition (\ref{effective.b}) leads to
an indeterminate value of $\beta _{\omega }$:
\begin{equation}
\beta _{\omega }\left( E\right) =-\frac{\partial }{\partial E}\ln \delta
\left( E-E_{0}\right) =indeterminate,  \label{iso.b.indeterminable}
\end{equation}%
which means that the thermostat inverse temperature $\beta _{\omega }$
admits \textit{any value} when the energy of the system is fixed at $E_{0}$.
This idea is schematically illustrated in FIG.\ref{gibbs.thermo.eps}.

\subsection{Fluctuations in the Gaussian approximation}

We shall suppose in the present analysis that the system and bath are large
enough in order to support a Gaussian approximation of the energy
fluctuations around the most probable macrostate $\bar{E}$. In addition, we
also assume that the energy dependence of the effective inverse temperature $%
\beta _{\omega }\left( E\right) $ allows for the existence of only one
intersection point $\bar{E}$ with the microcanonical caloric curve of the
system $\beta _{s}\left( E\right) $ in the thermal equilibrium condition (%
\ref{thermal.eq}). This latter requirement is just the condition of ensemble
equivalence. The average square dispersion of the internal energy $%
\left\langle \delta E^{2}\right\rangle $:
\begin{equation}
\left\langle \delta E^{2}\right\rangle =\int \left( E-\bar{E}\right)
^{2}\omega \left( E\right) \Omega \left( E\right) dE
\end{equation}%
can be estimated as follows:
\begin{equation}
\left\langle \delta E^{2}\right\rangle ^{-1}=-\frac{\partial ^{2}}{\partial
E^{2}}\left\{ \ln \omega \left( \bar{E}\right) +S\left( \bar{E}\right)
\right\} .  \label{aux}
\end{equation}%
The fluctuation $\delta \beta _{\omega }=\beta _{\omega }\left( E\right)
-\beta _{\omega }\left( \bar{E}\right) $ of the effective inverse
temperature can be related to the fluctuation $\delta E$ of the internal
energy of the system:
\begin{equation}
\delta \beta _{\omega }=-\frac{\partial ^{2}\ln \omega \left( \bar{E}\right)
}{\partial E^{2}}\delta E.  \label{aux2}
\end{equation}%
If we rewrite Eq.(\ref{aux}) in the following manner,
\begin{equation}
-\frac{\partial ^{2}\ln \omega \left( \bar{E}\right) }{\partial E^{2}}%
=\left\langle \delta E^{2}\right\rangle ^{-1}+\frac{\partial ^{2}S\left(
\bar{E}\right) }{\partial E^{2}},  \label{aux3}
\end{equation}%
and combine Eq.(\ref{aux2}) and (\ref{aux3}), we arrive at the correlation
between the effective inverse temperatures of the generalized thermostat and
the internal energy of the system as follows:
\begin{equation}
\left\langle \delta \beta _{\omega }\delta E\right\rangle =1+\frac{\partial
^{2}S\left( \bar{E}\right) }{\partial E^{2}}\left\langle \delta
E^{2}\right\rangle .  \label{tur}
\end{equation}%
This latter identity is a generalized expression of the fluctuation relation
(\ref{fluct1}). This fact can be noticed by rephrasing (\ref{tur}) as
follows:%
\begin{equation}
C=\beta ^{2}\left\langle \delta E^{2}\right\rangle +C\left\langle \delta
\beta _{\omega }\delta E\right\rangle  \label{tur2}
\end{equation}%
by using the microcanonical definition of the heat capacity, Eq.(\ref%
{micro.cap}), being $\beta =\beta _{s}\left( \bar{E}\right) $.

\section{Discussions}

We remark that the fluctuation relation (\ref{tur}) accounts for the
specific mathematical form of the probabilistic weight $\omega \left(
E\right) $ in an implicit way throughout the effective inverse temperature $%
\beta _{\omega }$. Once more, this result evidences the fact that the
system-surrounding energy interchange is effectively controlled by $\beta
_{\omega }$. The imposition of the restriction $\delta \beta _{\omega }=0$
associated with the Gibbs canonical ensemble (\ref{gibbs}) into Eq.(\ref%
{tur2}) leads to the usual identity $C=\beta ^{2}\left\langle \delta
E^{2}\right\rangle $. However, the restriction $\delta \beta _{\omega }=0$
is not compatible with the existence of energetic regions with negative heat
capacities $C<0$. Since the microcanonical entropy is locally convex $%
\partial ^{2}S\left( \bar{E}\right) /\partial E^{2}>0$ in such anomalous
regions, the identity (\ref{tur}) leads here to the inequality:
\begin{equation}
\left\langle \delta \beta _{\omega }\delta E\right\rangle >1.  \label{unc}
\end{equation}%
This means that any attempt to impose the canonical condition $\delta \beta
_{\omega }\rightarrow 0$, within regions with $C<0$, leads to the occurrence
of very large energy fluctuations $\delta E\rightarrow \infty $, and
conversely, any attempt to impose the microcanonical condition $\delta
E\rightarrow 0$ is accompanied by very large fluctuations of the effective
inverse temperature of the system surrounding $\delta \beta _{\omega
}\rightarrow \infty $. Remarkably, this behavior suggests the existence of
some kind of \textit{complementarity }between the internal energy $E$ and
the effective inverse temperature $\beta _{\omega }$ of the surrounding,
which is quite analogous to the complementarity between a coordinate $q$ and
its conjugated momentum $p$ in Quantum Mechanics! As well, the divergence $%
\delta \beta _{\omega }\rightarrow \infty $ when $\delta E\rightarrow 0$ is
not only applicable to regions where $C<0$: any attempt to reduce the energy
fluctuations to zero, $\delta E\rightarrow 0$, in Eq.(\ref{tur}) for any
fixed $\bar{E}$ leads to the following result:
\begin{equation}
\lim_{\delta E\rightarrow 0}\left\langle \delta \beta _{\omega }\delta
E\right\rangle =1,
\end{equation}%
indicating thus the divergence of the effective inverse temperature
fluctuations $\delta \beta _{\omega }\rightarrow \infty $ at this limit. The
present results allow to conclude:

\begin{description}
\item[\textbf{(i)}] \label{c1} While macrostates with $C>0$ are accessible
within the canonical ensemble, where takes place the restriction $\delta
\beta _{\omega }=0$, the anomalous macrostates where $C<0$ can be only
accessed when $\delta \beta _{\omega }\not=0$, that is, by using a
generalized thermostat whose energy dependence of its effective inverse
temperature $\beta _{\omega }\left( E\right) $ ensures the validity of the
inequality (\ref{unc}).

\item[\textbf{(ii)}] \label{c2} The total energy $E$ of the system and the
effective inverse temperature of the generalized thermostat $\beta _{\omega
} $ behave as \textit{complementary thermodynamic quantities} within the
regions with $C<0$.

\item[\textbf{(iii)}] \label{c3} The imposition of the microcanonical
restriction $\delta E\rightarrow 0$ leads at any internal energy $E$ to an
indetermination of the effective inverse temperature of the system
surrounding $\delta \beta _{\omega }\rightarrow \infty $ as a consequence of
the suppression of the underlying energetic interchange.
\end{description}

A deeper understanding of the physical meaning of the previous conclusions
is reached by discussing their implications on the two standard ways where
the external influence of the surrounding on the thermodynamic state of the
system is used within the Thermodynamics:

\begin{itemize}
\item as a \textit{control apparatus} (thermostat), or

\item as a \textit{measure apparatus} (thermometer).
\end{itemize}

\subsection{Implications on the thermodynamic control\label{control.m}}

It is well-known that the (inverse) temperature is, \textit{in general}, a
good control parameter for the internal energy of a system: the contact of a
thermostat with a given constant value $\beta $ leads to the existence of
small fluctuations of the internal energy $\delta E\propto \sqrt{N}$, where $%
N$ is the system size. The remarkable exception is that during the
first-order phase transitions. Here, a small variation on $\beta $ is able
to provoke a sudden change in the expectation value of the internal energy $%
\left\langle E\right\rangle $ of the system due to the multimodal character
of the energy distribution function in the neighborhood of the critical
inverse temperature $\beta _{c}$, which provokes very large energy
fluctuations $\delta E\propto N$. This physical situation was already
illustrated in FIG.\ref{potts.eps} of the introductory section for the case
of the thermodynamical description of the $q=10$ states Potts model on the
square lattice (\ref{potts}), where the origin of this anomaly relies on the
existence of inaccessible or unstable energetic regions with $C<0$.

From the perspective of the thermodynamic control, \textit{the fluctuation
relation of }Eq.(\ref{tur}) \textit{describes the necessary conditions where
the external influence imposed by the surrounding ensures the thermodynamic
stability of the system, allowing thus an effective control of the internal
energy} $E$ \textit{of the controlled system}. The term "effective control"
means that the internal energy of the system is preserved until the
precision of small fluctuations around the average value $\bar{E}$. This
kind of fluctuating behavior is ensured by the ensemble equivalence or the
existence of only one sharp peak in the energy distribution function $\rho
\left( E\right) dE$, which is mathematically expressed by the existence of
only one intersection point $\bar{E}$ in the condition of thermal
equilibrium (\ref{thermal.eq}).

\begin{figure}[tbp]
\begin{center}
\includegraphics[width=4.0413in]{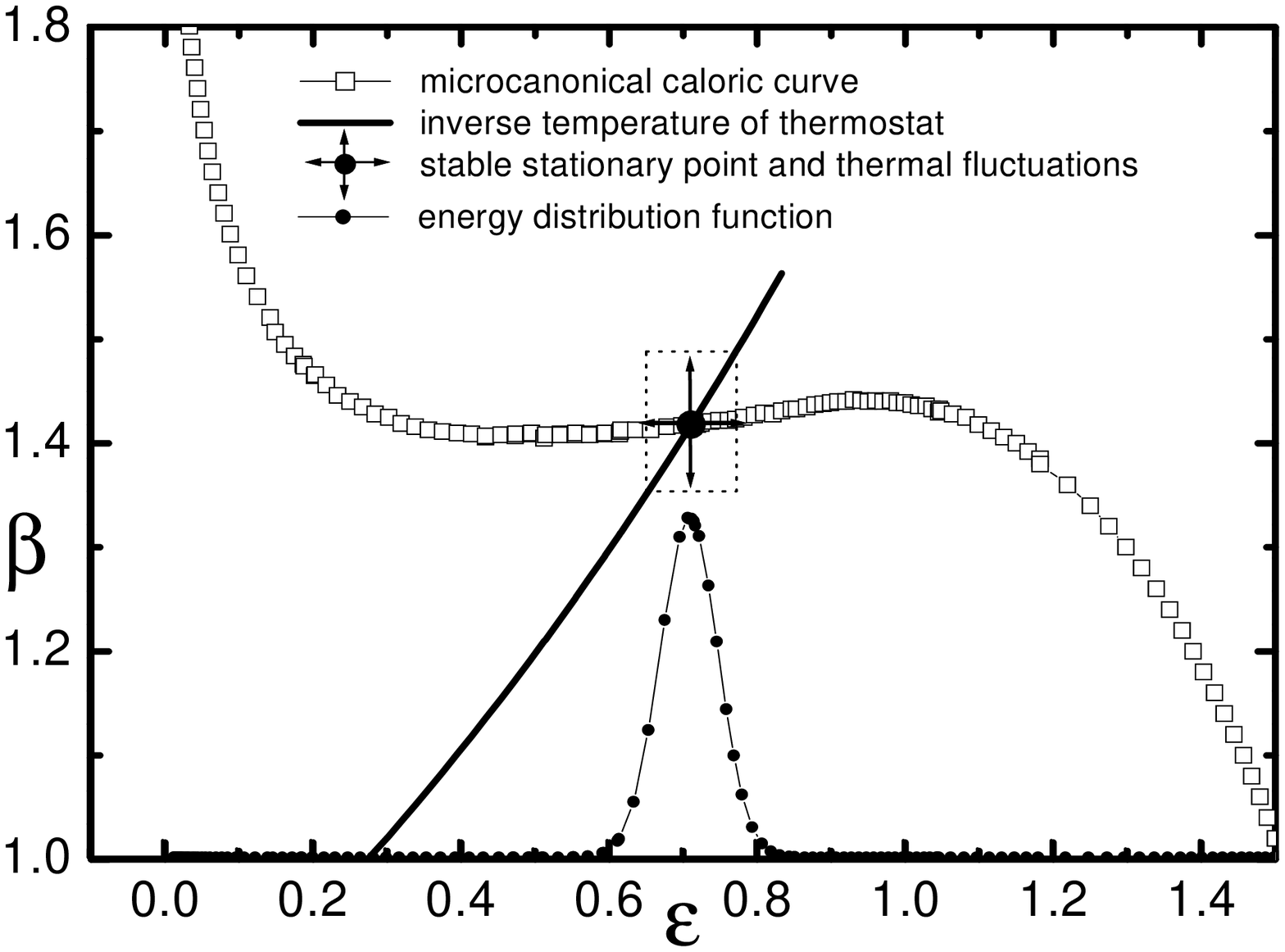}
\end{center}
\caption{Metropolis Monte Carlo calculations for the 10-state Potts model by
using Eq.(\protect\ref{interchange}). Consideration of a thermostat with a
variable inverse temperature $\protect\beta _{\protect\omega }\left(
E\right) $ can ensure the existence of a unique stable stationary point of
the microcanonical caloric curve $\protect\beta _{s}\left( E\right) $ where $%
\protect\beta _{\protect\omega }=\protect\beta _{s}$, in this way the system
can be forced to access its anomalous macrostates with $C<0$, and at the
same time, to eliminate the bimodal character of the energy distribution
function in the neighborhood of the critical inverse temperature $\protect%
\beta _{c}$.}
\label{gencanonical.eps}
\end{figure}

The Conclusion \textbf{(i)} claims basically that macrostates with $C<0$ can
be forced to be thermodynamically stable or accessible by using a
generalized thermostat with an appropriate energy dependence in its
effective inverse temperature $\beta _{\omega }\left( E\right) $. This
situation is clearly illustrated in FIG.\ref{gencanonical.eps}, where this
particular external influence has automatically eliminated the bimodal
character of the energy distribution function of the $q=10$ states Potts
model in the neighborhood of the critical temperature shown in FIG.\ref%
{potts.eps}.

The divergence $\delta E\rightarrow \infty $ when $\delta \beta _{\omega
}\rightarrow 0$ within the region with $C<0$ indicated in the Conclusion
\textbf{(ii)} accounts for the well-known fact that such macrostates turn
unstable or inaccessible within the canonical description. This divergence
is just a consequence of the Gaussian approximation which has been employed
to obtain the fluctuation relation (\ref{tur}), since the energy
fluctuations are actually on the order of the system size, $\delta E\propto
N $, which diverge only in the thermodynamic limit $N\rightarrow \infty $.
The Conclusion \textbf{(iii)} is just the mathematical result previously
obtained in Eq.(\ref{iso.b.indeterminable}) and illustrated in FIG.\ref%
{gibbs.thermo.eps}. Conveniently, the practical implications of this result
will be discussed in the next subsection.

As already shown, the consideration of a generalized thermostat with an
appropriate energy dependence of its effective inverse temperature $\beta
_{\omega }\left( E\right) $ is a simple but an effective consideration to
overcome the difficulties associated with the existence of macrostates with $%
C<0$. In principle, the practical implementation of the present ideas should
lead to the development of some experimental techniques which could be
particularly useful to deal with thermodynamical description of mesoscopic
systems which are now of interest of Nanosciences, and where the presence of
macrostates with $C<0$ is not unusual.

Another important framework of applications is to develop more efficient
Monte Carlo methods to deal with the difficulties associated to the presence
of first-order phase transitions \cite{mc3}. A simple and general way to
account for the presence of a generalized thermostat with a given effective
inverse temperature $\beta _{\omega }\left( E\right) $ is by using the
Metropolis method based on the acceptance probability of Eq.(\ref%
{interchange}). In fact, the microcanonical caloric curve $\beta _{s}\left(
\varepsilon \right) $ of the $q=10$ states Potts model shown in FIG.\ref%
{potts.eps} and FIG.\ref{gencanonical.eps} were obtained using this
methodology by calculating the averages $\left\langle \varepsilon
\right\rangle $ and $\left\langle \beta _{\omega }\right\rangle $, where the
validity of the Gaussian approximation for $N$ large enough and the
condition of thermal equilibrium (\ref{thermal.eq}) ensure the applicability
of the relations:
\begin{equation}
\bar{\varepsilon}=\left\langle \varepsilon \right\rangle ~~and~~\beta
_{s}\left( \bar{\varepsilon}\right) =\left\langle \beta _{\omega
}\right\rangle .  \label{metodo}
\end{equation}%
Numerical results derived from this algorithm agree very well with the ones
obtained using other Monte Carlo methods \cite{groff}.

\begin{figure}[tbp]
\begin{center}
\includegraphics[width=4.0075in]{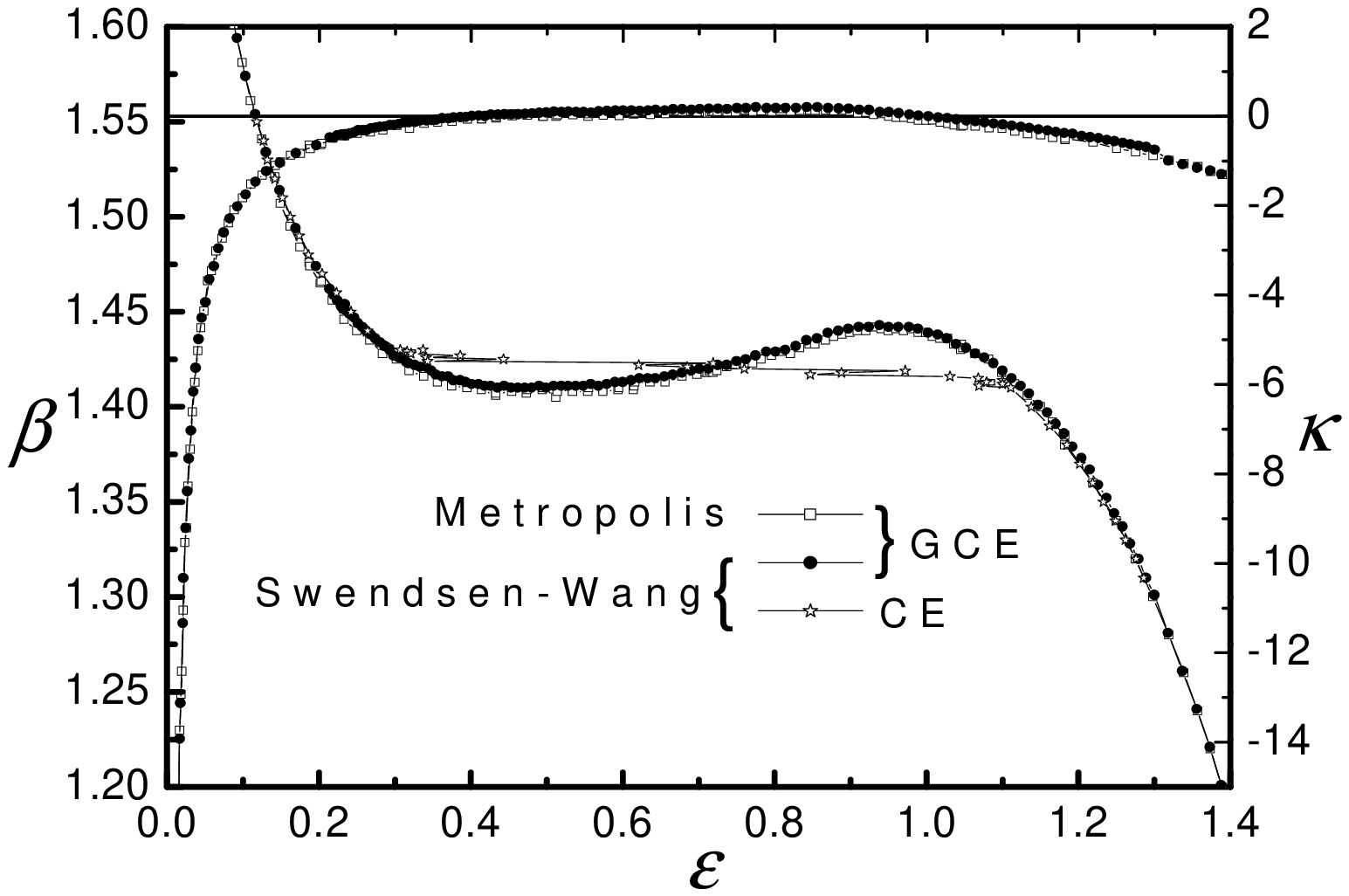}
\end{center}
\caption{The ordinary Swendsen-Wang cluster algorithm (CE) is unable to
account for the $C<0$ regime of the microcanonical caloric curve $\protect%
\beta _{s}\left( \protect\varepsilon \right) =\partial s\left( \protect%
\varepsilon \right) /\partial \protect\varepsilon $\ of the $q=10$ states
Potts model on the square lattice as shown here by our Monte Carlo
calculations. This difficulty is overcome by the Metropolis algorithm and
the Swendsen-Wang cluster algorithm based on the use of an effective inverse
temperature $\protect\beta _{\protect\omega }$ (GCE), where one can note a
remarkable agreement between these algorithms. In addition, we also show
here the energy per particle $\protect\varepsilon $ dependence of the second
derivative of the entropy per particle, $\protect\kappa \left( \protect%
\varepsilon \right) =\partial ^{2}s\left( \protect\varepsilon \right)
/\partial \protect\varepsilon ^{2}$, which has been obtained from the
fluctuation relation (\protect\ref{tur}) by calculating the fluctuations
quantities $\left\langle \protect\delta \protect\varepsilon %
^{2}\right\rangle $ and $\left\langle \protect\delta \protect\beta _{\protect%
\omega }\protect\delta \protect\varepsilon \right\rangle $. This dependence
clearly evidence the existence of an anomalous region where $\protect\kappa %
\left( \protect\varepsilon \right) >0\Leftrightarrow C<0$.}
\label{compara.eps}
\end{figure}

The idea of using the generalized thermostat with an appropriate effective
inverse temperature $\beta _{\omega }\left( E\right) $ can be easily
extended to other Monte Carlo methods based on the Gibbs canonical ensemble.
A specific example is the enhancement of the well-known Swendsen-Wang (SW)
cluster algorithm \cite{wolf}, which suffers from the \textit{supercritical
slowing down}\footnote{%
An exponential divergence of the correlation time $\tau _{N}$ in the
thermodynamic limit $N\rightarrow \infty $, $\tau _{N}\propto \exp \left(
N\right) $.} in its application to the $q=10$ states Potts model on the
square lattice and is unable to capture the $C<0$ regime of the
microcanonical caloric curve. The thermostat inverse temperature $\beta $
enters into this method by the probability for the cluster formation $%
p\left( \beta \right) =1-\exp \left( -\beta \right) $, which is used to
generate a new system configuration $X$. While the parameter $\beta $
remains constant in the original SW algorithm, our modification consists in
the substitution of this parameter by the effective inverse temperature of
the generalized thermostat $\beta _{\omega }$, which is redefined in each
Monte Carlo step, $\beta _{\omega }^{i}\rightarrow \beta _{\omega }^{i+1}$.
The parameter $\beta _{\omega }^{i+1}$ used to generate the configuration $%
X^{i+1}$ takes the value of the effective inverse temperature corresponding
to the total energy $E_{i}=H_{N}\left( X^{i}\right) $ of the previous system
configuration $X^{i}$, $\beta _{\omega }^{i}=\beta _{\omega }\left(
E_{i}\right) $. Here, $H_{N}$ is the Hamiltonian of Eq.(\ref{potts}) and $%
X=\left\{ \sigma _{k}\right\} $ the spin variables. The microcanonical
caloric curve is determined by using the same relations (\ref{metodo}) of
the Metropolis method explained before. In FIG.\ref{compara.eps} a
comparative study among the ordinary SW method (CE) and the Metropolis and
SW methods with an appropriate effective inverse temperature (GCE) is
depicted. The agreement between these last Monte Carlo algorithms and their
advantages over the first method is remarkable. As clearly evidenced, the
present ideas support the development of an alternative framework of the
well-known multicanonical Monte Carlo methods \cite{mc3}. As in FIG.\ref%
{gibbs.thermo.eps} as Monte Carlo calculations shown in FIG.\ref%
{gencanonical.eps} and FIG.\ref{compara.eps}, the following effective
inverse temperature:
\begin{equation}
\beta _{\omega }\left( E\right) =\eta \exp \left[ \lambda \left(
E-E_{0}\right) /N\right]
\end{equation}%
was assumed, where $\eta $ and $\lambda $ are two real positive parameters
controlling the horizontal position and the inclination of this dependence
respectively. The value $\lambda =0$ corresponds to the canonical ensemble (%
\ref{gibbs}), while $\lambda \rightarrow +\infty $ corresponds to the
microcanonical ensemble (\ref{micro}). More details of these calculations
can be found in the ref.\cite{vel2}.

\subsection{Implications on the thermodynamic measurements: An uncertainty
relation?}

Bohr and Heisenberg suggested in the past that the thermodynamic quantities
of temperature and energy are complementary in the same way as the position
and the momentum\ are related in Quantum Mechanics \cite{bohr}. Roughly
speaking their idea was that a \textit{definite temperature} can be
attributed to a system only if it is submerged in a heat bath (Gibbs
thermostat). Energy fluctuations are unavoidable. On the contrary, a
definite energy can be only assigned to systems in thermal isolation, thus
excluding the simultaneous determination of its temperature. Dimensional
considerations suggest the existence of the following relation:
\begin{equation}
\delta \beta \delta E\geq 1.  \label{b.unc}
\end{equation}%
However, these ideas have not reached a general consensus in the literature
\cite{Uffink}.One objection is that the mathematical structure of Quantum
theories is radically different from that of classical physical theories, so
that, there are no noncommuting observables in Thermodynamics.
Interestingly, the result expressed in Eq.(\ref{unc}) is quite analogous to
the uncertainty relation (\ref{b.unc}).

Before analyzing the implications of the fluctuation relation (\ref{tur}) in
the thermodynamic measurements, it is important to revise the concept of
temperature of a system. In the Bohr and Heisenberg arguments explained
above, as well as in the works of some other authors like the Landau (see
last paragraph of pag. 343 in ref.\cite{Landau}) and Mandelbrot \cite%
{MandelT}, \textit{a system has a definite temperature when it is put into
contact with a Gibbs thermostat}. In this viewpoint, the concept of the
temperature of an isolated system is unclear. We think this is a
misunderstanding, since the inverse temperature $\beta $ that enters in the
Gibbs canonical ensemble (\ref{gibbs}) is just the microcanonical inverse
temperature of the Gibbs thermostat (th), $\beta =\partial S_{th}/\partial
E_{th}$. This quantity does not properly characterize the thermodynamic
state of the system, but instead it characterizes the thermodynamic state of
the thermostat and its thermal influence on the system. Hereafter, \textit{%
we refer to the inverse temperature of a system defined as its
microcanonical inverse temperature} $\beta _{s}=\partial S\left( E\right)
/\partial E$. From this perspective, the temperature of an isolated system
is a well-defined quantity and does not undergo fluctuations, $\delta
E=0\Rightarrow \delta \beta _{s}=0$.

Although it is possible to obtain the system temperature at a given energy
by calculating the Boltzmann entropy $S\left( E\right) $, the practical
determination of the temperature is always imprecise. The energy is a
quantity with a \textit{mechanical significance}, and it can be determined
by performing only one instantaneous measurement on an isolated system. The
temperature is a quantity with a \textit{thermo-statistical significance},
whose determination demands to appeal to the concept of statistical
ensemble. For example, it is derived from a statistics of measurements or
temporal averages of certain physical observables with a mechanical
significance usually referred to \textit{thermometric quantities} (e.g.: the
length of a mercury column, an electric signal, the pressure of an ideal
gas, the average form of a particles distribution function, etc.). In
practice, the temperature $\beta _{s}$ of a given system\ is\textit{\
indirectly} \textit{measured} from its interaction with another system
(usually smaller than the system under study and referred to thermometer
(th)), whose internal dependence of its temperature $\beta _{th}$ on a
thermometric quantity is \textit{previously known}. The fundamental key
supporting this procedure is precisely the condition of thermal equilibrium:
\begin{equation}
\beta _{s}=\beta _{th}.  \label{thermal.eq.2}
\end{equation}

In our approach, the surrounding can be also used as a measurement apparatus
(thermometer), so that, the quantity $\beta _{th}$ is just the effective
inverse temperature $\beta _{\omega }$. It is possible to realize that the
generalized thermostat used in the previous Monte Carlo calculations has the
dual role: to be a control and a measure apparatus at the same time. In
fact, the energy dependence\ of the microcanonical inverse temperature $%
\beta _{s}$ of the Potts model is \textit{a priori} unknown, and, we
estimate it through the effective inverse temperature of the generalized
thermostat $\beta _{\omega }$ via the condition of thermal equilibrium in
Eq.(\ref{metodo}). This kind of computational procedure exhibits essential
features of a real determination of the energy-temperature dependence
(caloric curve).

Despite the apparent simplicity, the determination of the energy-temperature
dependence of a system by using this procedure is rather complex, which is
evidenced from the two factors affecting its precision:

\begin{itemize}
\item[\textbf{A. }] This measurement process necessarily involves a \textit{%
perturbation on the thermodynamic state of the system}.

\item[\textbf{B. }] The interaction also affects the thermodynamic state of
the thermometer in an uncontrollable stochastic way, which provokes the
existence of \textit{errors in the determination of the effective inverse
temperature} $\beta _{\omega }$ used to estimate $\beta _{s}$ via the
condition of thermal equilibrium.
\end{itemize}

It is possible to see in the Conclusions \textbf{(ii)} and \textbf{(iii)}
that the precision factors \textbf{A} and \textbf{B} are rather \textit{%
complementary}. The error type \textbf{A} can be characterized in terms of
the system energy fluctuations $\delta E$, since the inverse temperature
fluctuation $\delta \beta _{s}$ is directly correlated with $\delta E$. The
error type \textbf{B} is characterized in terms of the effective inverse
temperature fluctuations $\delta \beta _{\omega }$, which may also depend on
$\delta E$, but in an indirect way. According to the Conclusion \textbf{(iii)%
}, any attempt to reduce the perturbation of the system to zero, $\delta
E\rightarrow 0$, (the error type \textbf{A}) leads to a progressive
increasing of the error type \textbf{B}, $\delta \beta _{\omega }\rightarrow
\infty $, which affects the estimation of $\beta _{s}$ by using $\beta
_{\omega }$. The error type \textbf{B} can reduce to zero $\delta \beta
_{\omega }\rightarrow 0$ by imposing the conditions of the Gibbs canonical
ensemble (\ref{gibbs}), which always involves certain perturbations of the
system energy $\delta E\not=0$ (error type \textbf{A}). This perturbation is
relatively small when the system size is large enough and the heat capacity
of the system is positive $C>0$. However, this situation changes radically
when the macrostate of the system is characterized by a negative heat
capacity $C<0$. According to the Conclusion \textbf{(ii)}, the reduction of
the error type \textbf{B} to zero, $\delta \beta _{\omega }\rightarrow 0$,
induces the thermodynamic instability of the macrostates with $C<0$ and
leads to the existence of very large energy fluctuations $\delta
E\rightarrow \infty $. In practice, we should admit the simultaneous
existence of the errors type \textbf{A} and \textbf{B}, which can reasonably
be small and unimportant, when the sizes of the system under study and the
thermometer are large enough. However, such errors are much significant when
the system size $N$ is small (in a system of few bodies or constituents)
that the concept of system temperature becomes experimental unobservable,
and therefore, physically meaningless.

As already evidenced, \textit{the fluctuation relation of }Eq.(\ref{tur})
\textit{accounts in some way for the limit of precision for a determination
of the energy-temperature dependence of a given system by using a
measurement procedure based on thermal equilibrium with another system }%
(thermometer). Although this qualitative behavior is quite close to the Bohr
and Heisenberg intuitive idea of energy-temperature complementarity, the
fluctuation relation (\ref{tur}) does not always support a complementary
relationship between the system energy $E$ and the effective inverse
temperature $\beta _{\omega }$ or the system energy $E$ and inverse
temperature $\beta _{s}$. The reason is that the mathematical structure of
Eq.(\ref{tur}) does not have the form of a complementary relation.
Fortunately, this limitation is not difficult to overcome.

\subsection{Generalization: The Quantum-Statistical Mechanics analogy}

The derivation of the restricted fundamental result (\ref{tur}) relies on
the Gaussian approximation. However, there is a simple way to overcome this
difficulty. The inverse temperature fluctuation $\delta \beta _{s}$ can be
expressed for small $\delta E$ in the following way:
\begin{equation}
\delta \beta _{s}=\frac{\partial ^{2}S}{\partial E^{2}}\delta E;
\end{equation}%
therefore, the fluctuation relation (\ref{tur}) can be rephrased as follows:%
\begin{equation}
\left\langle \delta \eta \delta E\right\rangle =1,  \label{dif}
\end{equation}%
where $\eta $ is the difference between the effective inverse temperature of
the generalized thermostat and the microcanonical inverse temperature of the
system:%
\begin{equation}
\eta =\beta _{\omega }-\beta _{s}.
\end{equation}

The validity of result (\ref{dif}) does not depend on the Gaussian
approximation. In order to show this, let us denote by $\rho \left( E\right)
=\omega \left( E\right) \Omega \left( E\right) $ the energy distribution
function of the ansatz (\ref{new}). The distribution function $\rho \left(
E\right) $ is not arbitrary. It obeys the following general mathematical
conditions:

\begin{itemize}
\item[C1. ] \textit{Existence:} The distribution function $\rho \left(
E\right) $ is a nonnegative, bounded, and differentiable function on the set
$\pi \subset R$ of all physically admissible energies $E$.

\item[C2. ] \textit{Normalization:} This function obeys the normalization
condition:%
\begin{equation}
\int_{\pi }\rho \left( E\right) dE=1.
\end{equation}

\item[C3. ] \textit{Boundary conditions:} This function vanishes together
with its first derivative on boundary $\partial \pi $ of the set $\pi $:%
\begin{equation}
\forall E\in \partial \pi :\rho \left( E\right) =\frac{\partial }{\partial E}%
\rho \left( E\right) =0.
\end{equation}
\end{itemize}

The key of the demonstration is the consideration of the following identity:%
\begin{equation}
\left\{ -\frac{\partial }{\partial E}\right\} \rho \left( E\right) =\eta
\rho \left( E\right) ,  \label{identity}
\end{equation}%
which allows to associate the quantity $\eta $ to an "operator" $\hat{\eta}$:%
\begin{equation}
\hat{\eta}=-\frac{\partial }{\partial E}.
\end{equation}%
By using the identity (\ref{identity}) as well as the properties (C2) and
(C3)\ of the distribution function $\rho \left( E\right) $, it is easy to
obtain the following results:
\begin{equation}
\left\langle \eta \right\rangle =\int_{\pi }\eta \rho \left( E\right)
dE=\int_{\pi }\hat{\eta}\rho \left( E\right) dE=0,  \label{thermal.gen}
\end{equation}%
\begin{equation}
\left\langle E\eta \right\rangle =\int_{\pi }E\eta \rho \left( E\right)
dE=\int_{\pi }E\hat{\eta}\rho \left( E\right) dE=1.  \label{correlation.gen}
\end{equation}%
The result (\ref{thermal.gen}) is just the condition of thermal equilibrium,
Eq.(\ref{thermal.eq}). The fundamental result (\ref{dif}) is immediately
obtained from Eq.(\ref{correlation.gen}) by operating as follows:
\begin{equation}
\left\langle \eta \right\rangle =0\Rightarrow \left\langle E\eta
\right\rangle -\left\langle E\right\rangle \left\langle \eta \right\rangle
\equiv \left\langle \delta E\delta \eta \right\rangle =1.
\end{equation}%
The result (\ref{dif}) can be rephrased one more time by using the
well-known \textit{Schwarz inequality}:
\begin{equation}
\left\langle \delta A^{2}\right\rangle \left\langle \delta
B^{2}\right\rangle \geq \left\langle \delta A\delta B\right\rangle ^{2}
\end{equation}%
arriving finally at a definitive result:

\begin{equation}
\Delta \eta \Delta E\geq 1,  \label{def.tur}
\end{equation}%
where $\Delta \eta =\sqrt{\left\langle \delta \eta ^{2}\right\rangle }$ and $%
\Delta E=\sqrt{\left\langle \delta E^{2}\right\rangle }$.

Undoubtedly, the expression (\ref{def.tur}) has the form of "an uncertainty
relation" that exhibits now a very general validity. It clarifies that the
\textit{complementary relation actually exists for the system energy} $E$
\textit{and the inverse temperature difference} $\eta $ \textit{between the
system and its surrounding} (acting as a thermostat or a thermometer). Any
attempt to perform an exact determination of the system temperature $\beta
_{s}$ throughout the thermal equilibrium, $\Delta \eta \rightarrow 0$,
involves a strong perturbation on the system energy $\Delta E\rightarrow
\infty $, thus, $E$ becomes indeterminable; any attempt to reduce this
perturbation to zero, $\Delta E\rightarrow 0$, makes impossible to determine
the system inverse temperature $\beta _{s}$ by using the condition of
thermal equilibrium since $\Delta \eta \rightarrow \infty $.

Remarkably, the above thermodynamic complementarity between $E$ and $\eta $
is quite analogous to the complementarity in Quantum theories. Contrary to
what was preliminarily objected, this complementarity could be also
supported in terms of the noncommutativity of mathematical operators $\hat{E}%
\equiv E$ and $\hat{\eta}\equiv -\partial _{E}$:
\begin{equation}
\left[ \hat{E},\hat{\eta}\right] =1\Leftrightarrow \Delta \eta \Delta E\geq
1.  \label{SM}
\end{equation}%
Thus, the relations of Eq.(\ref{SM}) can be considered the Statistical
Mechanics counterpart of the familiar quantum relations:
\begin{equation}
\left[ \hat{q},\hat{p}\right] =i\Leftrightarrow \Delta q\Delta p\geq 1,
\label{QM}
\end{equation}%
where $\hslash \equiv 1$. The formal correspondence is: $q\sim E$ and $p\sim
\eta $. This Quantum-Statistical Mechanics analogy can be also extended to
the properties of the distribution functions: $\rho \left( q\right)
=\left\vert \psi \left( q\right) \right\vert ^{2}$ and $\rho \left( E\right)
=\omega \left( E\right) \Omega \left( E\right) $, since the quantum
distribution function $\rho \left( q\right) $ also obeys the properties
(C1), (C2) and (C3). Demanding a little of imagination, the behavior of the
energy distribution function in the neighborhood of the critical point
illustrated in FIG.\ref{potts.eps} possesses a certain analogy with the
tunneling of the distribution function $\rho \left( q\right) $ (a wave
packet) throughout a classical barrier; that is, the phase coexistence
phenomenon can be interpreted as a "tunneling" of the energy distribution
function $\rho \left( E\right) $ throughout the canonically inaccessible
region.

Obviously, such a Quantum-Statistical Mechanics analogy should not be
overestimated, although it is licit to recognize that it is very
interesting: (1) Both are physical theories with a statistical mathematical
apparatus; (2) While Quantum Mechanics is a theory hallmarked by the
Ondulatory-Corpuscular dualism, Statistical Mechanics is also hallmarked by
another kind of dualism since it works simultaneously with physical
quantities with a purely \textit{mechanical significance} (e.g.: energy) and
physical quantities with a purely \textit{thermo-statistical significance}
(e.g.: inverse temperature); (3) Thermodynamics is the counterpart theory of
Classical Mechanics: while Classical Mechanics assumes the simultaneous
determination of position $q$ and momentum $p$ when $\hslash \rightarrow 0$,
Thermodynamics assumes the simultaneous determination of the system energy $%
E $ and its temperature $\beta _{s}$ in the thermodynamic limit $%
1/N\rightarrow 0$.

Our approach to uncertainty relations in Thermodynamics constitutes an
improvement of the works of Rosenfeld and Sch\"{o}lgl \cite%
{Rosenfeld,Scholgl}, which have been also based on \textit{Fluctuation theory%
} \cite{Landau}. These authors derived their respective formulations from
the consideration of a surrounding in the thermodynamics limit, and hence,
this physical situation actually corresponds to the Gibbs canonical ensemble
or in general the Boltzmann-Gibbs distribution \cite{Uffink}. Since this
work hypothesis presupposes a concrete thermodynamic influence, \textit{an
experimentalist has no free will to change the fluctuational behavior of the
system in a given thermodynamic state}.

This important limitation is overcome in our work: the fluctuational
behavior of the system-surrounding equilibrium is modified by considering a
different energy dependence of the effective inverse temperature $\beta
_{\omega }\left( E\right) $, which presupposes to use a different
experimental arrangement. In addition, it is not necessary to appeal to
\textit{Statistical Inference theory} to arrive at an uncertainty relation,
as in the Mandelbrot development \cite{Mandelbrot}, which undergoes the same
"free will limitation" of the Rosenfeld and Sch\"{o}lgl works explained
above since it is applicable only to the framework of the Gibbs canonical
ensemble. Interestingly, no one of the above approaches says anything about
the relevance of anomalous macrostates with $C<0$ on the practical
determination of the system energy-temperature dependence.

\section{Conclusions}

In an attempt to establish an appropriate framework in order to deal with
the existence of macrostates with negative heat capacities in terms of a
fluctuation theory, we have introduced the concept of effective inverse
temperature $\beta _{\omega }$, which characterizes the thermodynamic
influence of the surrounding on the system under study, and specially,
allows to extend the condition of thermal equilibrium to a wide class of
equilibrium or meta-equilibrium situations. Precisely in terms of this
quantity we arrive at a suitable generalization of the well-known canonical
relation between the heat capacity and the energy fluctuation, $C=\beta
^{2}\left\langle \delta E^{2}\right\rangle $. The new identity, Eq.(\ref{tur}%
), defines a criterion capable of detecting macrostates with negative heat
capacities of the system under study through the correlated fluctuations of
the effective inverse temperature $\beta _{\omega }$ of the surrounding
(generalized thermostat) and the energy $E$\ of the system itself. This last
constitutes a novel kind of mechanics to perform a more effective
thermodynamic control on those anomalous macrostates hidden behind the
phenomenon of ensemble inequivalence, which inspires the developments of new
experimental techniques of control to deal with the thermodynamic
description for mesoscopic systems,\ like those which are now of interest in
Nanosciences, as well as the introduction of new Monte Carlo methods to
overcome the difficulties associated with the presence of first-order phase
transitions.

We think that a significant aspect of this paper is to provide new physical
arguments supporting the existence of a complementary relation, between the
system energy and temperature, as postulated by Bohr and Heisenberg in the
early days of the Quantum Mechanics. Both the fluctuation relation (\ref{tur}%
) and its generalization (\ref{def.tur}), which can be regarded as an
uncertainty relation, impose limitations to the precise determination of the
energy-temperature dependence (caloric curve) of the system under study by
using a measurement procedure based on thermal equilibrium with another
system\textit{\ }(thermometer). While this limitation is unimportant in
large enough systems, it discards the practical utility of some concepts of
Thermodynamics in the context of systems with a small number of
constituents. Surprisingly, these results suggest the existence of a
remarkable analogy between the Statistical Mechanics and the Quantum
Mechanics.

\section*{Acknowledgments}

It is a pleasure to acknowledge partial financial support by FONDECYT
3080003 and 1051075. L.V. also thanks the partial financial support by the
project PNCB-16/2004 of the Cuban National Programme of Basic Sciences.

\end{document}